\documentclass{aa}
\usepackage{amsmath,graphicx,amssymb}
\usepackage{txfonts}
\usepackage{natbib}
\bibpunct{(}{)}{;}{a}{}{,}
\def\urltilda{\kern -.15em\lower .7ex\hbox{\~{}}\kern .04em}
\newcommand\vir{\object{CU\,Vir}}
\newcommand\ori{\object{V901\,Ori}}
\newcommand{\zav}[1]{\left(#1\right)}

\newlength\staretab

\makeatletter
\def\sgn{\mathop{\operator@font sgn}\nolimits}
\makeatother

\begin{document}

\title{Surprising variations in the rotation of the chemically peculiar
    stars CU Virginis and V901 Orionis}
\titlerunning{Surprising rotation variations of the mCP stars CU Vir and V901 Ori}

\author{Z.~Mikul\'a\v sek\inst{1,2}
    \and J.~Krti\v{c}ka\inst{1}
    \and G.\,W.~Henry\inst{3}
    \and J.~Jan\'{i}k\inst{1}
    \and J.~Zverko\inst{4}
    \and J.~\v{Z}i\v{z}\v{n}ovsk\'{y}\inst{5}
    \and M.~Zejda\inst{1}
    \and J.~Li\v{s}ka\inst{1}
    \and P.~Zv\v{e}\v{r}ina\inst{1}
    \and D.\,O.~Kudrjavtsev\inst{6}
    \and I.\,I.~Romanyuk\inst{6}
    \and N.\,A.~Sokolov\inst{7}
    \and T.~L\"{u}ftinger\inst{8}
    \and C.~Trigilio\inst{9}
    \and C.~Neiner\inst{10}
    \and S.\,N.~de Villiers\inst{11} }
\authorrunning{Z.~Mikul\'a\v sek et al.}
\offprints{Zden\v ek~Mikul\'a\v sek,\\
\email{mikulas@physics.muni.cz}}

\institute{Department of Theoretical Physics and Astrophysics,
           Masaryk University, Kotl\'a\v{r}sk\'a 2, CZ\,611\,37, Brno, Czech Republic
        \and Observatory and Planetarium of Johann Palisa, V\v SB --
            Technical University, Ostrava, Czech Republic
        \and Center of Excellence in Information Systems, Tennessee
            State University, Nashville, Tennessee, USA
        \and Tatransk\'a Lomnica 133, SK\,059\,60, Slovak Republic
        \and Astronomical Institute of Slovak Academy of Science,
            Tatransk\'{a} Lomnica, Slovak Republic
        \and Special Astrophysical Observatory of RAS, Nizhnij Arkhyz,
            Russia
        \and Central Astronomical Observatory at Pulkovo, Russia
        \and Institute for Astronomy of the University of Vienna,
            Vienna, Austria
        \and NAF - Osservatorio Astrofisico di Catania, Italy
        \and LESIA, Observatoire de Paris, CNRS, UPMC, Universit\'e Paris
            Diderot; 5 place Jules Janssen, 92190 Meudon Cedex, France
        \and  Private Observatory, 61 Dick Burton Road, Plumstead,
            Cape Town, South Africa}

\date{Received 29 July 2011/ Accepted 20 September 2011}

\abstract{The majority of magnetic chemically peculiar (mCP) stars
exhibit periodic light, radio, spectroscopic and spectropolarimetric
variations that can be adequately explained by the model of a rigidly
rotating main-sequence star with persistent surface structures. CU Vir
and V901 Ori belong among these few mCP stars whose rotation periods
vary on timescales of decades.} {We aim to study the stability of the
periods in CU Vir and V901 Ori using all accessible observational data
containing phase information.} {We collected all available relevant
archived observations supplemented with our new measurements of these
stars and analysed the period variations of the stars using a novel
method that allows for the combination of data of diverse sorts.} {We
found that the shapes of their phase curves were constant during the
last several decades, while the periods were changing. At the same
time, both stars exhibit alternating intervals of rotational braking
and acceleration. The rotation period of CU Vir was gradually
shortening until the year 1968, when it reached its local minimum of
0.52067198 d. The period then started increasing, reaching its local
maximum of 0.5207163 d in the year 2005. Since that time the rotation
has begun to accelerate again. We also found much smaller period
changes in CU Vir on a timescale of several years. The rotation period
of V901 Ori was increasing for the past quarter-century, reaching a
maximum of 1.538771 d in the year 2003, when the rotation period began
to decrease.} {We propose that dynamical interactions between a thin,
outer magnetically-confined envelope, braked by the stellar wind, with
an inner, faster rotating stellar body is able to explain the observed
rotational variability. A theoretically unexpected alternating
variability of rotation periods in these stars would remove the
spin-down time paradox and brings a new insight into structure and
evolution of magnetic upper-main-sequence stars.}

\keywords {stars: chemically peculiar -- stars: variables -- stars:
individual: CU~Vir, V901~Ori -- stars: rotation}

\maketitle

\section{Introduction}

In addition to mass, all stars inherit a fraction of the angular
momentum of the mother cloud out of which they condensed.
Consequently, all stars rotate. Stars spend the prevailing part of
their lives as main-sequence (MS) objects fusing hydrogen in their
cores. During the MS phase, the angular momentum of stars without
mass-loss is conserved. This implies that stellar rotation period
changes should occur on the timescale of the moment of inertia
variations, which is of the order of $10^7 - 10^9$ years
\citep{meynet}. How do we test this?

The magnetic chemically peculiar (mCP) stars, which have an abnormal
surface chemical composition, are the most suitable test beds for
studying the rotational evolution in upper MS stars. The overabundant
elements in their atmospheres are concentrated into large, persistent
spot regions. As the star rotates, periodic variations in the
brightness, spectrum, and magnetic field are observed. Combining both
present and archival observations of mCP stars collected over the past
several decades, we can reconstruct their rotational evolution with
unprecedented accuracy.

Careful period analyses of photometric observations of dozens of the
best-monitored mCP stars listed in the On-line catalogue of
photometric observations of magnetic chemically peculiar stars
\citep{mcpod} have been performed in \citet{unstead}. They confirm our
theoretical expectations that the rotation periods and light curves of
most upper MS stars are constant on timescales of decades.  A few mCP
stars display secular cyclic changes in the shape of their light
curves \citep[see e.g.][]{zigasx}, which can be attributed to the
precession of magnetically distorted stars \citep{pyper04}. However,
there is also a small subgroup of hot mCP stars that have stable light
curve shapes and spectroscopic variability, but exhibit variable
rotation periods \citep{pyper98,mik901,town}. Constancy of their light
curves on the scale of decades excludes precession as the cause of the
observed period changes \citep[for details see][]{mik901}.

This letter concentrates on the period analysis of the best-monitored
mCP stars -- CU Virginis and V901 Orionis.

\section{The stars}

\subsection{CU Virginis}

\vir\ = HD~124224 = HR~5313 is one of the most enigmatic stars in the
upper MS. This very fast rotating silicon mCP star has a moderate,
nearly dipolar magnetic field \citep{bolan,pyper98} with the pole
strength of $B_{\rm{p}}=3.0$~kG tilted towards the rotational axis by
$\beta=74^{\circ}$, the axis inclination being $i\simeq43^{\circ}$
\citep{trigi00}. \vir\ is the only known MS star that shows variable
radio emission, resembling the radio lighthouse of pulsars
\citep{trigi,trigi11,ravi}, it also displays variations in light and
spectral lines of \ion{He}{i}, \ion{Si}{ii}, \ion{H}{i}, and other
ions. The nature of its variability is minutely studied in
\citet{krtcu}. \vir\ belongs to the most frequently studied mCP stars.

Occasional rapid increases in its rotation period have been reported
and discussed. \citet{pyper97,pyper98} discovered an abrupt increase
of the period from $0\fd5206778$ to $0\fd52070854$ that occurred
approximately in 1984 and \citet{pyper04} discussed two possible
scenarios of the explanation of the observed O-C diagram, namely a
continually changing period or two constant periods. After the year
1998 \citet{trigi,trigi11} observed another increase in the period of
radio pulses of $\Delta P=1.12$~s with respect to the period
determined by \citet{pyper98}. However, the 2010 measurements indicate
some period decrease. The rate of the deceleration can be evaluated
using the spin-down time, $\tau$, defined as $\tau=P/\bar{\dot{P}}$,
where $P(t)$ is the instantaneous rotation period at the time $t$ and
$\bar{\dot{P}}$ is the mean rate of the rotational deceleration. The
paradox of \vir\ is according to \citet{unstead} that its spin down
time, $ \tau \sim 6\times 10^5$ years, is more than two orders of
magnitude shorter than the estimated age of the star -- $9\times10^7$
yr \citep{kobacu}.

\subsection{V901 Orionis}

V901 Ori = HD 37776 is a young hot star (B2\,IV) residing in the
emission nebula IC 432, with a global, extraordinarily strong
$(B_s\approx20$~kG), and unusually complex magnetic field
\citep{thomla,ko901}. The observed moderate light variations are
caused by the spots of overabundant silicon and helium \citep{krt901}.

Three decades of precise photometric and spectroscopic monitoring have
revealed a continuous rotational deceleration \citep{mik901},
increasing the period of about $1\fd5387$ by a remarkable 18 s! Ruling
out (a) the light-time effect in a binary star, (b) the precession of
the star's rotational axis, and (c) evolutionary effects as possible
causes of the period change, we interpreted the deceleration in terms
of the rotational braking of the outer stellar layers as caused by the
angular momentum loss in the stellar magnetosphere.

However, this cannot explain the discrepancy between the spin-down
time, $\tau = 2.5 \times 10^5$ yr, and the star's age of one million
years or older \citep{mik901,mik7355,unstead,kobacu}. The
interpretation of the rotation period evolution by simple angular
motion loss was also questioned by the negative value of the second
derivative of the period: $\ddot{P}=-29(13)\times
10^{-13}\,\mathrm{d}^{-1}$, which indicated that braking could change
into acceleration \citep{mik901}. Nevertheless, the significance of
this conclusion was fairly low.

\begin{figure}[t]
\centering
\resizebox{1.\hsize}{!}{\includegraphics{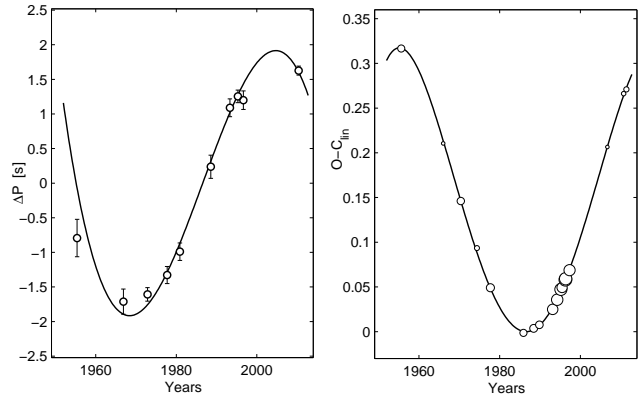}}
\caption{Long-term variations of the rotation period and O-C values
for the magnetic chemically peculiar star CU Vir. Left: The model
curve of long-term rotational period changes $\Delta P$ of \vir\ in
seconds with respect to the mean period $P_0 = 0\fd52069415$. The
formal accuracy of the polynomial fit is comparable with the thickness
of the fitted line. Right: The difference of the observed and
calculated times of zero phase according to a linear ephemeris in
days. Each point represents the average of 498 consecutive individual
measurements; the weights of these means are indicated by their
areas.} \label{Figcu}
\end{figure}

\section{Observations}

For the period analyses of \vir\ we obtained some new photometric or
spectroscopic data, specifically: 1)~JK derived 210 individual
spectrophotometric magnitudes in 21 bands (201 - 299 nm) by processing
10 IUE spectra, in 1979 January - March  \citep[for details
see][]{krtcu};\quad 2)~GWH obtained 374 precise $\mathit{BV}$
measurements at Fairborn Observatory Arizona, USA, in 2010 February -
2011 June;\quad 3)~JL obtained 38 $V$ measurements using his own
microtelescope at the SAAO, South Africa and Brno, CR, in 2010 April -
June ;\quad 4)~JJ obtained 251 $\mathit{vb}$ observations at Suhora
Observatory, Poland, in 2011 March; and JJ+JL obtained 536
$\textit{UBV}$ measurements using the 0.5m reflector of the SAAO,
South Africa, in 2011 May.

We collected a total of 8965 individual measurements of \vir\ obtained
between 1949 and 2011 (62 years or 43662 revolutions of the star)
including 8270 photometric and spectrophotometric measurements in
photometric bands from 200 to 753 nm as well as spectroscopic,
spectropolarimetric and radiometric observations. The diverse
observational data originated from 39 sources (see the list in
Appendix B).

The \ori\ data listed in \citet{mik901} were enhanced with 98 precise
$BV$ measurements obtained by GWH at Fairborn Observatory Arizona, USA
from September~2009 to February~2011. We used a total of 2611
individual measurements obtained from 1976 to 2011 from 14 sources. In
addition to 2409 photometric measurements, the data include 202
measurements of equivalent widths of eleven selected \ion{He}{i}
lines.

\section{The models and their results}

The method of data processing, outlined in the Appendix A, assumes
that phase curves of all monitored quantities are constant, while the
`instant period' of changes $P(t)$ is variable in time $t$. It is
advantageous to introduce a `phase function'
$\vartheta(t,\mathbf{b})$, which is a continuous monotonic function of
time $t$. The fractional part of it corresponds to the common phase
$\varphi$, the integer part is the so-called epoch $(E)$ and
$\mathbf{b}$ is a set of $g_{\mathrm{b}}$ free parameters describing a
model of the period development. Quantities $P(t)$ and
$\vartheta(t,\mathbf{b})$ are bound by the simple equality
$P(t)=1/\dot{\vartheta}$, where the dot refers to the first time
derivative of the quantity. Because we did not know the true nature of
observed period changes of $P(t)$ we chose standard low-orders
polynomials for their description.

\subsection{CU Virginis} \label{cuvir}

The period changes $P(t)$ (see Fig.\,\ref{Figcu}) can be adequately
approximated by a cubic parabola with the origin at the time $t=T_0$,
\begin{equation}\label{cuper}
P(t)=P_0+A\zav{3\mathit{\Theta}-4\,\mathit {\Theta}^3};\quad
\mathit{\Theta}=\zav{t-T_0}/\mathit{\Pi},
\end{equation}
where $A$ and $\mathit{\Pi}$ are parameters expressing the amplitude
and timescale of the observed period changes, and $\mathit{\Theta}$ is
a time-like function. The instant period $P(t)$ reaches its local
extrema $P_0\mp A$ at
$\mathit{\Theta}_{1,2}=\mp\textstyle{\frac{1}{2}}$, where $P_0$ is the
period at the time of $t(\mathit{\Theta}=0)=T_0$. The value
$\mathit{\Pi}$ then equals the duration of the rotation deceleration
epoch that took place in the time interval
$t_{1,2}=T_0\mp\textstyle{\frac{1}{2}}\mathit{\Pi}$.

Applying the equality $\dot{\vartheta}=1/P(t)$, we obtain the
following approximation for the phase function in the form of the
fourth-order polynomial of time
\begin{equation}\label{cuteta}
\vartheta(t,\textbf{b})\cong\vartheta_0-\frac{B}{P_0} \zav
{\textstyle{\frac{3}{2}}\mathit{\Theta}^2-\mathit {\Theta}^4};\
\vartheta_0=\frac{t-M_0}{P_0},\  B=\frac{A\,\mathit{\Pi}}{P_0},
\end{equation}
where $\vartheta_0$ is the phase function for a linear ephemeris with
the origin at $M_0 \equiv 2\,446\,730.4447$ and the basic period
$P_0$.

Using the above formulated phenomenological model, we derive $P_0 =
0\fd52069415(8)$, $\mathit{\Pi} = 13260(70)\,\mathrm{d} = 36.29(19)$
yr, $T_0 = 2\,446\,636(24)$ or 1986.56(7), $A = 1.915(10)$~s, and
$B=0.5643(29)$~d. The period $P(t)$ reached its local minimum in the
year 1968.4, $P_{\mathrm{min}}= 0\fd52067198(7)$, and its local
maximum in 2004.7: $P_{\mathrm{max}}= 0\fd52071631(18)$. The
rotational deceleration rate reached its maximum
$\dot{P}=0.158$\,s\,yr$^{-1}$ in the year 1986.6. The zero phase times
 of \vir\ can be evaluated using the relation
$\textit{JD}(k)\cong
M_0+P_0\,k+B\,(\textstyle{\frac{3}{2}}\mathit{\Theta_k}^2-\mathit
{\Theta_k}^4)$, where $k$ is an integer and
$\mathit{\Theta}_k=(M_0+P_0\,k-T_0)/\mathit{\Pi}$.

\begin{figure}[t]
\centering
\resizebox{0.80\hsize}{!}{\includegraphics{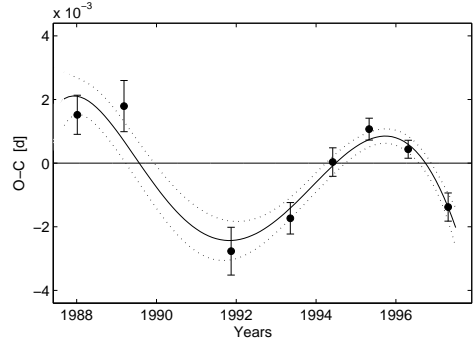}}
\caption{Medium-term variations in the rotational period of \vir. The
modulation of the long-term rotational changes in the observed time of
the light minimum minus the calculated time of light minimum is
expressed in days. Owing to uneven sampling within the observing
seasons, we were not able to detect rotational modulation on shorter
timescales.} \label{Figdetail}
\end{figure}

The basic model fits the observed long-term period changes very well
and enables us to predict the zero phase times with an accuracy of
0.001 d. However, the analysis of the residuals from the accepted
polynomial model reveals an additional variation. Between 1988 and
1998, for which there is excellent photometric coverage
\citep{ade92,pyper98}, we find a small modulation of the period on a
timescale of several years (Fig.\,\ref{Figdetail}).

\subsection{V901 Orionis}

For the phase function of \ori\ we adopted polynomials in the form of
the third-order Taylor-expansion, also used in \citet{mik901}:
\begin{align} \label{901teta}
&\vartheta(t,\mathbf{b})=\vartheta_0-\textstyle{\frac{1}{2}}\dot{P}_0
\vartheta_0^2-\textstyle{\frac{1}{6}}P_0\, \ddot{P}\vartheta_0^3;\quad
\vartheta_0=(t-M_0)/P_0;\nonumber\\ &P(t)=1/\dot{\vartheta}\cong
P_0\zav{1+\dot{P}_0\vartheta_0+\textstyle{\frac{1}{2}}P_0
\ddot{P}\vartheta_0^2},
\end{align}
where $\vartheta_0$ is the phase function for the linear ephemeris
with the origin at $M_0$. $P_0$ and $\dot{P}_0$ are the instantaneous
period and its first derivative at $M_0$. We assumed that the second
derivative of the period, $\ddot{P}$, is constant throughout the
interval of observations.

\begin{figure}[t]
\centering
\resizebox{0.95\hsize}{!}{\includegraphics{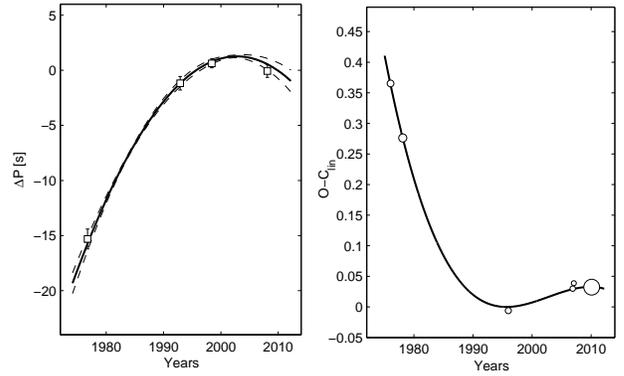}}
\caption{Long-term variations in the rotational period and O-C diagram
of \ori. Left: Variation in the rotation period $\Delta P$ in seconds
with respect to the mean period $P_0 = 1\fd538756(3)$, derived from
2611 individual measurements. The 1-$\sigma$ deflections from the
quadratic fit are denoted by dashed lines. Points with error bars
correspond to virtual period deflections for appropriately selected
groups of consecutive individual data items. Right: The difference of
the observed zero phase time and the time calculated according to the
linear ephemeris in fractions of a day. Each point represents an
average of 435 consecutive measurements; the weights of these means
are indicated by their areas.} \label{Fig901}
\end{figure}

We put the origin, $M_0 = 2\,449\,967.969(5)$, at a brightness maximum
(in 1999). Our fitted parameter values include $P_0 = 1\fd538756(3)$,
the instantaneous period at time $M_0$, and $\dot{P}_0 =
11.3(0.7)\times 10^{-9} = 0.356(23)$\,s\,yr$^{-1}$ and
$\ddot{P}=-4.4(6)\times
10^{-12}\mathrm{d}^{-1}=-1.38(19)\times10^{-5}$\,s\,yr$^{-2}$, the
first and second time derivatives. The negative value of $\ddot{P}$ is
now established with 7-$\sigma$ certainty. Accordingly, the star
reached its longest period, $P_{\mathrm{max}}= 1\fd5387709(15)$, in
$2002.8\pm1.1$, and is now accelerating again (see
Fig.\,\ref{Fig901}). The zero-phase timing can be calculated by the
relation $\textit{JD}(k)\cong
M_0+P_0\,(k+\textstyle{\frac{1}{2}}\dot{P}_0\,k^2+
\textstyle{\frac{1}{6}}P_0\ddot{P}\,k^3)$, where $k$ is an integer.
There is no indication of any short-term period modulation analogous
to that seen in \vir.

\section{Discussion}

We were able to reveal unexpected alternating lengthenings and
shortenings in the periods of \vir\ and \ori\ owing to simultaneous
processing of all available observational material obtained during
many decades, including our own observations made in 2007--11. The
period evolutions of both stars were modelled by polynomials of low
degrees. In the case of \vir\ we also tested a fit with a harmonic
function \citep{unstead} and a Gaussian function corresponding to the
transient nature of the phenomenon. The only differences among them
are noticeable at the very beginning of our observations, which
slightly favour the latter two alternatives.

The possible harmonic-like changes of the observed periods of both
stars together with the strict constancy of the phase curves can in
principle be explained by the light time effect caused by an invisible
companion orbiting the star in a nearly circular orbit. However, this
orbital motion would be revealed by considerable radial velocity
variations $\Delta\mathit{RV}$. In the case of \vir\ the observed
change in the period $\Delta P$ of 3.8~s corresponds to a variation
$\Delta \mathit{RV}=c \Delta P/P= 25$~km\,s$^{-1}$, while the observed
change in \ori\ period of $\Delta P\simeq 20$~s predicts a
$\Delta\mathit{RV}$ variation of even 45~km\,s$^{-1}$! However, no
such long-term $RV$ variations synchronised with period changes have
been found \citep{pyper97,mik901}. Additionally, the light time effect
cannot explain the complex medium-term period variations observed in
\vir\ (see Sec.~\ref{cuvir}, Fig.\,\ref{Figdetail}).

We conclude that the observed period variations are caused by uneven
rotation of surface layers of the stars \citep{step,mik901,unstead}.
However, the physical explanation of the observed rotation period
variations is not straightforward. Rotational braking by angular
momentum (AM) loss via a magnetised stellar wind \citep{mik901} cannot
be the sole cause of the period variations because we detected
intervals when the rotational period decreases. Consequently, the AM
loss may affect only the outer stellar envelope that is hardened by
the global magnetic field, leaving the faster rotating core
unaffected. Intervals of rotational deceleration may then alternate
with intervals of angular momentum exchange between the slowly
rotating envelope and the faster rotating core. The latter manifests
itself by the rotational acceleration of the outer envelope.

However, this explanation is theoretically challenging. Given the
intermediate age of \vir, the detected rotation variations are likely
not connected with the transient behaviour that is connected with
settling into the MS stage. Therefore, we can expect a stable inner
configuration in both stars. As follows from \citet{flora} and
\citet{brano}, the magnetic fields confined to the outer stellar
layers only are unstable. On the other hand, a magnetic field that
penetrates deep into a star causes nearly uniform rotation
\citep{moss,uhlaci,mame}.

The possible role of stellar winds in the mechanism of the rotational
instability is supported by the fact that stronger period variations
are detected in \ori, which is more luminous than \vir\ and,
consequently, has a stronger stellar wind \citep{krku}. Moreover, this
could help to explain the fact that the younger star \ori\ rotates
more slowly than \vir.

All these considerations should be the subject of additional physical
modelling. Nevertheless, the aim of this letter is more unpretentious
-- we only want to draw the attention to an interesting property of
the rotation of some upper MS stars.

\section{Conclusions}

Our study investigated the nature of the rotational period variations
in two well-observed mCP stars \vir\ and \ori. Contrary to the results
of previous studies, we show that the period changes are not monotonic
-- intervals of rotational deceleration alternate with intervals of
rotational acceleration, all on the timescale of several decades.
These results explain the spin-down time paradox of these stars,
according to which the spin-down time was significantly shorter than
the age of these stars. On the other hand, this unexpected behaviour
of two fairly dissimilar mCP stars poses a strong challenge for any
theoretical models.

\begin{acknowledgements}
We thank R. von Unge, D.\,M. Pyper, M. Lenc, S.\,J. Adelman, T.
Ryabchikova, and R. Arlt for valuable discussions and unpublished
observational data. This work was funded by the following grants: GAAV
IAA301630901, GA\v{C}R 205/08/H005, MEB 061014/WTZ CZ 10-2010, MEB
0810095, VEGA 2/0074/09, and MUNI/A/0968/2009. We also thank the
referee D.\,Moss for valuable comments and inspiring suggestions.
\end{acknowledgements}

\newpage
\appendix

\section{The outline of the method}\label{virtoc}

The techniques used to analyse the data are based on the rigourous
application of non-linear, weighted, least-squares methods used
simultaneously for all relevant data that contain phase information.
Our technique does not use an O-C diagram as an intermediate stage of
data processing; O-C diagrams are used only as a visual check on the
adequacy of the models. Let us assume that all observed \textit{phase
curves} of a star are adequately described by the unique general model
function $F(\vartheta,\mathbf{a})$, described here by $g_a$ parameters
contained in a parameter vector $\mathbf{a}$,
$\mathbf{a}=(a_1,...,a_j,...,a_{g_a})$ . In our computation we assume
that the form of all the phase variations is constant and that the
time variability of the observed quantities are given by a
\textit{phase function} $\vartheta(t,\mathbf{b})$, which is a
monotonic function of time $t$. The fractional part of it corresponds
to the common phase, the integer part is the so-called epoch $(E)$. We
can express the phase function by means of a simple model quantified
by $g_b$ parameters $\mathbf{b}$,
$\mathbf{b}=(b_1,...,b_k,...,b_{g_b})$. The instantaneous period is
given simply by the equality $P(t)=1/\dot{\vartheta}$.

For a realistic modelling of phase variations for all data types for a
star, we need $g_a$ free parameters for the description of the model
function $F(\vartheta,\mathbf{a})$ and $g_b$ free parameters for the
description of the phase function $\vartheta(t,\mathbf{b})$ . The
computation of the free parameters was iterative under the basic
condition that the weighted sum $S(\mathbf{a,b})$ of the quadrates of
the difference $\Delta y_i$ of the observed value $y_i$ and its model
prediction is minimal ($w_i$ being the individual weight of the $i$-th
measurement).
\begin{align}\label{LSM}
&\Delta y_i=y_i-F(\vartheta_i);\quad S=\sum^n_{i=1}\Delta y_i^2
w_i;\quad
\delta S=\mathbf{0};\quad \Rightarrow \\
&\sum^n_{i=1}\Delta y_i\ \frac{\partial
F(\vartheta_i,\mathbf{a})}{\partial a_j}\ w_i=0;\quad
\sum^n_{i=1}\Delta y_i\ \frac{\partial F}{\partial \vartheta_i}\
\frac{\partial \vartheta(t_i,\mathbf{b})}{\partial b_k}\
w_i=0\nonumber
\end{align}

We obtain here $g = g_a + g_b$ equations of $g$ unknown parameters.
The weights of individual measurements $w_i$ are inversely
proportional to their expected uncertainty. The system is non-linear;
we have to determine the parameters iteratively. With a good initial
estimate of the parameter vectors $a$ and $b$, the iterations converge
very quickly. Usually we need only several tens of iterations to
complete the iteration procedure.

\subsection{Virtual O-C diagrams. Evolution of periods.}

The short-term modulation of the phase function was analysed by means
of the residuals of the observed data $\Delta y_i$, creating
individual values of the phase shifts expressed in days (O-C)$_j$ with
adapted individual weight $W_j$ for each observed datum and averages
of the phase shifts defined for arbitrarily selected groups of
measurements $\overline{\textrm{(O-C)}}_i$ or deflection of the mean
period from the instant model period $\Delta P_k(t_k)$:
\begin{align}
\textrm{(O-C)}_j&=-P(t_j)\,\Delta y_j \zav{\frac{\partial F}{\partial
\vartheta}}^{-1};\quad W_j=\zav{\frac{\partial F}{\partial
\vartheta}}^2\,w_j;\\
\overline{\textrm{(O-C)}}_k&=\frac{\sum_{j=1}^{n_k}\,(\textrm{O-C})_j\,
W_j}{\sum_{j=1}^{n_k}\,W_j};\quad \Delta P_k=
\frac{\sum_{j=1}^{n_k}\,(\textrm{O-C})_j\,\vartheta_j\,
W_j}{\sum_{j=1}^{n_k}\,\vartheta_j^2\,W_j}. \nonumber
\end{align}\label{virtual}

Computations of $\overline{\textrm{(O-C)}}_k$ and $\Delta P_k$
followed after the model parameters were found; consequently they had
no influence on the model solution. They were used only to visualise
of the solution. Similarly, we can compute virtual `observed' values
of the instant period from a group of observations to generate the
model curves in our figures.

Relation \ref{virtual} can also be used to determine reliably
 zero-phase times for selected groups of observations
O$_k$. These quantities depend only marginally  on the chosen model of
the phase function $\vartheta(t,\mathbf{b})$. Therefore we can use
these O$_k$ in the process of the phase function modelling.

\section{Brief specification of \vir\ data}

\begin{table}[ht]
\begin{center}
\caption{Data with phase information about \vir\ used in the
 period analysis.}
\begin{tabular}{llrl}
  \hline
season/s & Data type & $N$ &  Reference \\
\hline
1949-52 & EW \ion{He}{i}  & 34 & \citet{deutsch} \\
1955 & $\mathit{UBV}$ & 162 & \citet{hardie} \\
1964-6 & $\mathit{UBV}$ & 702 & \citet{abul}\\
1964 & EW \ion{Si}{ii} & 40 & \citet{peter} \\
1964 & EW \ion{He}{i}  & 9 & \citet{peter} \\
1965-6 & RV H$\gamma$ & 16 & \citet{abt}\\
1966 & $B$ & 33 & \citet{hardie+}\\
1967,76 & EW \ion{He}{i}  & 37 & \citet{hard}  \\
1967,76 & EW \ion{Si}{ii}  & 19 & \citet{hard}  \\
1968-9 & EW \ion{Si}{ii} & 124 & \citet{krivo} \\
1968-9 & EW H$\gamma$ & 21 & \citet{krivo}  \\
1968 & $\mathit{UBV}$ & 287 & \citet{blanco}\\
1972-81 & spf mag & 179 & \citet{pypad}\\
1972 & $\mathit{UBV}$ & 29 & \citet{win}\\
1974 & ${\beta}$ & 95 & \citet{weiss} \\
1974 & $\mathit{uvbyW}$ & 460 & \citet{weiss}\\
1975 & spf OAO II & 54 & \citet{molnar} \\
1976-8 & $B_{\rm{eff}}$ & 14 & \citet{bolan}\\
1977 & EW \ion{He}{i}  & 69 & \citet{peder}\\
1979 & spf IUE & 210 & \citet{krtcu}  \\
1980-3 & $\mathit{uvby}$ & 89 & \citet{pypad}\\
1980-3 & $\beta$ & 23 & \citet{pypad}\\
1983 & EW \ion{He}{i} & 9 & \citet{hies} \\
1987-9 & $\mathit{UBV}$ & 1012 & \citet{ade92}\\
1988 & $\beta$ & 25 & \citet{mus}\\
1990-3 & $B_{\rm{T}}V_{\rm{T}}H_{\rm{p}}$ &  288 & \citet{esa98} \\
1991-7 & $\mathit{uvby}$ & 3635 &  \citet{pyper98}\\
1994-5 & EW \ion{He}{i}  & 19 & \citet{kusch}\\
1995 & RV H$\delta$ & 19 & \citet{kusch} \\
1995 & EW \ion{Si}{ii}  & 19 & \citet{kusch}\\
1995 & EW H$\delta$ & 18 & \citet{kusch} \\
1995 & $B_{\rm{eff}}$  & 21 & \citet{pyper98}\\
1998-09 & radio & 5 & \citet{ravi} \\
2002-9 & $V$ & 353 & \citet{pojm} \\
2010-11 & $BV$ & 374 & Henry - this paper \\
2010-11 & EW \ion{Si}{ii} & 59 & Jan\'ik - this paper \\
2010 & $V$ & 38 & Li\v{s}ka - this paper \\
2011 & $vb$ & 251 & Jan\'ik - this paper \\
2011 & $\mathit{UBV}$ & 402 & Jan\'ik \& Li\v{s}ka - this paper \\
  \hline
\end{tabular}\label{Vir}
\end{center}
\end{table}
\noindent
The data used for the analysis of \vir\ are given in
Table~\ref{Vir}. Here we used the following abbreviations: EW -- the
equivalent width, RV -- radial velocity, $\beta$ -- H$\beta$
photometry, $\mathit{UBV}$ -- Johnson $\mathit{UBV}$ photometry,
$\mathit{uvby}$ -- Stroemgren $\mathit{uvby}$ photometry,
$B_{\rm{T}}V_{\rm{T}}H_{\rm{p}}$ -- Hipparcos photometry,
$B_{\rm{eff}}$ -- the mean longitudinal magnetic induction,  spf --
magnitudes derived from spectrograms obtained by UV satellites OAO\,2
and IUE, and radio - timings of radiopulses.

\end{document}